# Matter-wave Beam Splitter on an Atom Chip for Portable Atom-Interferometer

S. J. Kim,[1] H. Yu,[1,2] S. T. Gang,[1] and J. B. Kim[1]

[1]*Department of Physics Education, Korea National University of Education, Chung-Buk 363-791, Republic of Korea*
*E-mail address: jbkim@knue.ac.kr*
[2]*Department of Physics and JILA, University of Colorado, Boulder, Colorado 80309-0440, USA*

We construct a matter-wave beam splitter using $^{87}$Rb Bose-Einstein condensate on an atom chip. Through the use of radio-frequency-induced double-well potentials, we were able to split a BEC into two clouds separated by distances ranging from 2.8 μm to 57 μm. Interference between these two freely expanding BECs has been observed. By varying the rf-field's amplitude, frequency, or polarization, we investigate behaviors of the beam-splitter. From the perspective of practical use, our BEC manipulation system is suitable for application to interferometry since it is compact and the repetition rate is high due to the anodic bonded atom chip on the vacuum cell. The portable system occupies a volume of 0.5 m$^3$ and operates at a repetition rate as high as ~0.2 Hz.

**PACS number(s):** 03.75.Be, 03.75.Lm, 39.10.+j, 39.90.+d

## I. INTRODUCTION

The wave nature of matter enables the construction of atom interferometry measuring the phase of atomic wave function [1]. Atom interferometry has led scientific advances in diverse fields, including fundamental quantum physics, precision metrology, and atomic physics [2]. Since the first experimental observation of interference of Bose-Einstein condensates (BECs) [3], intensive studies on confined-atom interferometers with BECs have been performed [4-18]. A common phase and the confinement of the atoms in a BEC offer long interrogation time for precision measurements. In addition, highly accurate control of the location of the atomic waves is essential in experiments for studying spatially varying fields or interactions [2].

Coherent splitting a BEC into a double-well is equivalent to an optical beam splitter, and acts as a key element in a confined matter-wave interferometer. The most prominent matter-wave beam splitters are constructed by using optical potentials [7, 19, 20] or rf-dressed potentials on an atom chip [9, 12]. The rf-dressed double-well potential can be achieved by rf-coupling of different spin states of an electronic ground state, and the vector character of the magnetic coupling allows precise control and versatility of the potentials. Applying multifrequency rf-fields can additionally enhance the freedom of the potentials [21]. In addition, with demand for realizing a miniaturized interferometer, integration with other matter-wave devices, and controlling quantum states on a microscale, atom chips have gained recognition as an ideal platform for the matter-wave beam splitter [22]. Even though atom chips have enormously simplified the making process of BECs and on-chip interferometry has advanced rapidly in recent decades, the systems for interferometry are still too large to be applied for use in the field [23]. From the perspective of wider applications, portability, repetition-rate, and energy efficiency of the system are crucial properties of the interferometer.

In this paper, we present a matter-wave beam splitter on an atom chip using a compact BEC manipulation system with a high-repetition-rate. The system occupies a volume of 0.5 m$^3$ and operates at a repetition rate as high as 0.2 Hz. Initially prepared $^{87}$Rb BEC in a magnetic single-well potential is split by deforming the trap into a double-well using rf-dressing on the trapped atoms [24]. Interference between the two freely expanding BECs has been observed. By adjusting the amplitude and the frequency of the rf-field, the separation distances have been controlled from ~2.8 μm to ~57 μm. In addition, we investigated the orientation of the double-well by varying the polarization of the rf-field.

## II. COMPACT BEC MANIPULATION SYSTEM

The experimental apparatus, details of which are given in [25, 26], is similar to that used in previously reported compact atom chip experiments [23, 27]. Fig. 1 shows the configuration of our commercial product vacuum chamber (RuBECi, ColdQuanta). To achieve high-repetition-rates and a low pressure, a typical dual-chamber system is adopted [23]. The upper and the lower chamber are separated by a silicon disk with a 750 μm diameter hole at the center. The lower chamber contains a high-pressure background vapor of $^{87}$Rb (~10$^{-8}$ Torr), provided by using a dispenser, while the upper chamber is pumped to maintain an ultrahigh vacuum state (< low 10$^{-10}$ Torr) by a 2 l/s ion pump for differential pumping.

In the 19 × 19 × 40 mm³ lower Pyrex cell, atoms are loaded into a two-dimensional (2D) magneto-optical trap (MOT) using the quadrupole field created by permanent magnets around the cell. An upward cooling beam acts as a pushing beam for the trapped atoms to go through the hole at the silicon disk and its reflection by the disk makes a 2D⁺ MOT configuration [28]. All experimental steps, laser cooling with a six-beam (3D MOT configuration), evaporative cooling, matter-wave beam splitting and detection take place in the 19 × 19 × 65 mm³ upper Pyrex cell where the cell's top is replaced by the atom chip. The atom chip is anodic bonded on the cell. Anodic bonding allows an ultrahigh vacuum by baking the chamber at high temperatures of up to 400 °C. The cell is surrounded by a 50 × 50 × 90 mm³ cage of coil pairs. A 500 × 500 mm² double-layer table holds the vacuum chamber and the optics for cooling and imaging.

Table I shows the typical timing sequence of a BEC preparation for the experiments. A 3D MOT is loaded with ~10⁹ atoms in ~1.2 s from the 2D⁺ MOT. Compressing the MOT followed by 6.2 ms optical molasses cools the atoms to <40 μK. The molasses is optically pumped into the $F = m_F = 2$ ground state and then loaded into a magnetic trap that is manipulated by an external Z-coil placed directly above the atom chip. By reducing the big Z-coil current, the trapped atoms are adiabatically moved toward near the chip's surface within 400 ms. Those atoms are transferred into the chip magnetic trap by ramping-off the external Z-coil while ramping-on the chip wire currents on the atom chip. The trap is then gradually compressed to perform rf-evaporation. The rf-field for the rf-evaporation is generated by driving a loop (diameter ~1 cm) attached on the bottom of the external Z-coil. After the 1st rf-evaporation step, a 100 ms decompression-ramp transfers the atoms to the final trap located 120 μm directly below the z-trap wire, with radial and axial trap frequencies of 1043 Hz and 42 Hz, respectively. The 2nd rf-evaporation step reduces the cloud temperature below the transition condition for BEC formation. The prepared BEC typically contains ~2 × 10⁴ atoms without recognizable thermal fraction.

Detection for a distribution of the atoms is performed via resonant absorption imaging on a CCD camera (Basler acA 2000-50gm NIR) with a 100 μs illumination time.

## III. MATTER-WAVE BEAM SPLITTER ON ATOM CHIP

To implement a double-well for constructing a matter-wave beam splitter on an atom chip, we adopt a rf-induced potential scheme [24, 29]. This scheme is based on a combination of static and rf-magnetic fields forming a dressed adiabatic potential resulting from rf-coupling of different spin states. Following [29], an effective rf-dressed adiabatic potential at the position $r$ can be written as

$$V_{ad}(r) = m_F g_F \mu_B \sqrt{\Delta(r)^2 + \Omega(r)^2} . \qquad (1)$$

Here $m_F$ is the magnetic quantum number of the state, $g_F$ is the Landé g-factor, and $\mu_B$ is the Bohr magneton. The detuning $\Delta(r)$ and the coupling term $\Omega(r)$ are given by:

$$\Delta(r) = |B_S(r)| - \frac{\hbar \omega}{|g_F \mu_B|} , \qquad (2)$$

$$\Omega(r) = \frac{B_{rf}(r) \times B_S(r)}{2|B_S(r)|} , \qquad (3)$$

where $B_S$ is the static trapping field, $B_{rf}$ is the rf-field, $\hbar$ is the reduced Planck constant, and $\omega$ is the angular frequency of the rf-field. Since the coupling term $\Omega(r)$ has dependency on the relative orientation of the static and the rf-field, inhomogeneous coupling strength around the static trap minima deforms a single-well into a double-well potential for linear polarization of the rf-field [30].

The matter-wave beam splitter is fully integrated on the atom chip. The atom chip has been described comprehensively elsewhere [26]. The 26 × 26 mm² sized atom chip has four deposited 10 μm-thick copper wires on its vacuum side surface. Fig. 2 (a) shows the central region of the atom chip for the atom interferometer. The central trap wire (z-shaped, width 100 μm) running DC current is used with external bias fields to create a static magnetic potential. DC current is run through the dimple wire (width 100 μm) transversely crossing the other three wires to enhance the longitudinal confinement. Consequently, atoms are confined in a Ioffe-Pritchard trap and the confining magnetic field $B_S$ can be written as

$$B_S(r) = G\rho \cos\phi\, e_x - G\rho \sin\phi\, e_y + B_I e_z , \qquad (4)$$

in the vicinity of the trapping potential minimum. Here $G$ is the gradient of the static trap, $B_I$ is the magnitude of the Ioffe field, and $\rho = \sqrt{x^2 + y^2}$ and $\phi = \arctan(y/x)$ in cylindrical coordinates. The rf-wires (width 50 μm) on each side separated by 102 μm from the trap wire provide an oscillating magnetic field $\boldsymbol{B}_{rf}$ that creates the adiabatic rf-induced potentials

$$\boldsymbol{B}_{rf} = B_{rf}^A \boldsymbol{e}_x \cos(\omega t) + B_{rf}^B \boldsymbol{e}_y \cos(\omega t - \delta). \quad (5)$$

Here $B_{rf}^A$ and $B_{rf}^B$ are the amplitudes of the rf-fields from rf-wire A and B, respectively, and $\delta$ is the relative phase shift. Since the trapped atoms are close to the atom chip surface, running rf-current through the atom chip wires with less than 200 mA is sufficient to generate strong rf-fields for deforming the static single-well into a rf-induced double-well potential.

The splitting distance of the double-well $d$ has been controlled over a wide range as a function of the amplitude and the frequency of the rf-field. With negative detuning $\omega < |g_F \mu_B| B_I / \hbar$, the geometry of the rf-induced potential is dominantly determined by the coupling term. Consequently, the two minima are separated by

$$d \simeq \sqrt{2\left(|\boldsymbol{B}_{rf}|^2 - B_C^2\right)/G}, \quad (6)$$

where $B_C^2 = 2B_I(B_I - \hbar\omega/g_F\mu_B)$, in the area of $\rho \ll B_I/G$ [29]. On the contrary, with positive detuning $\omega > |g_F\mu_B| B_I / \hbar$, the resonance term becomes dominant for the shape of the rf-induced potential and the separation can be approximated as

$$d \simeq \frac{2\hbar\omega}{g_F\mu_B G} \quad (7)$$

at the region far from the static potential minima. Figs. 3(a) and (b), which present the calculated rf-dressed double-well potentials based on the typical Ioffe-Pritchard static trap configuration, show the dependences of the split distance $d$ on the rf-amplitude and rf-frequency, respectively.

In addition to controllability of the split distances, the use of two rf-wires allows control over the polarization of the rf-field corresponding to the orientation of the rf-induced double-well. For the linear polarization, $\delta = 0$ or $\pi$, the coupling term of Eq. (3) becomes

$$\Omega(r)^2 = \frac{B_{rf}^{A2} + B_{rf}^{B2}}{8|\boldsymbol{B}_S(r)|^2}\left\{2B_I^2 + G^2\rho^2\left[1 - \cos(2\alpha)\cos(2\phi) \pm \sin(2\alpha)\sin(2\phi)\right]\right\}, \quad (8)$$

where $\alpha = \arctan(B_{rf}^B / B_{rf}^A)$ and the positive (negative) sign of the last term is for $\delta = 0$ ($\delta = \pi$). The rf-induced potentials have two minima at $\alpha$, $\alpha + \pi$ for $\delta = 0$ and at $-\alpha$, $-\alpha + \pi$ for $\delta = \pi$. Fig. 3(c) shows the typical cases of $B_{rf}^A = B_{rf}^B$ with $\delta = 0$ and $\pi$, which lead to the vertical and horizontal double-well, respectively.

*1. Small distance splitting: negative detune*

For small splitting distances (<6 μm), the amplitudes of the rf-currents through the rf-wires, $I_{rf}^A$ and $I_{rf}^B$ (for rf-wire A and B, respectively), are ramped linearly from zero to their final values, typically 70~80 mA. The Larmor frequency at the minima of the static trap is $g_F\mu_B B_I / h \simeq 700$ kHz, and the rf-frequency is fixed at 510 kHz to prevent rf-forced spin flip. With $B_{rf}^A = B_{rf}^B = B_{rf}$ (that is, $I_{rf}^A = I_{rf}^B = I_{rf}$) and $\delta = 0$, the single-well is deformed into a vertical double-well in 15 ms ramp time as illustrated in Fig. 2(a).

In the realistic situation of implementing a double-well, the different distances from the current carrying wires result in unwanted asymmetry on the rf-induced double-well. Furthermore, gravity affects the asymmetry of the double-well, in particular for the vertical splitting situation. To balance the split BECs by compensating the asymmetry, an additional rf-field along the z-direction is applied by running rf-current through the dimple wire with relatively small amplitudes (<3 mA) [31]. Fig. 4 shows the calculated typical double-well potentials with a realistic wire configuration including the gravity effect. Asymmetry of the vertical double-well potential is nearly compensated by the rf-current through the dimple wire with proper amplitude. This balancing scheme will be introduced in another paper and is outside the area of interest of this paper. Briefly, the compensation is achieved by modulating the coupling strength by tilting the polarization of the total rf-field.

In order to study the coherence of the splitting, we turn off the trapping fields and release the condensates to allow them to fall freely for a time $T = 15$ ms. Typical matter-wave interference patterns from the expanded and overlapped clouds are obtained by taking resonant absorption images with side imaging light, as shown in Fig. 2(b). Due to the effect of projecting the atoms trapped in the rf-dressed potential onto different Zeeman sublevels during the turn-off, the interference pattern was not clear. By applying a magnetic gradient field along the z-direction during time-of-flight, atoms in the different magnetic substates are spatially separated and clear interference patterns for each $m_F$ state are obtained (Fig. 2(c)). Since the fringe spacing $\Lambda$ is related to the initial separation of the split BECs $d$, we determine the fringe spacing by fitting a cosine function on a Gaussian envelope $G(y)\left[1+c\cos(2\pi y/\Lambda+\phi)\right]$ to the measured density profile. Fig. 5 shows (a) the interference patterns and (b) measured fringe spacings for different rf-amplitudes $I_{rf}$. Ref. [3] predicts the fringe spacing $\Lambda$ for an expanding non-interacting gas from two points with distance $d$ as

$$\Lambda = 2\pi\hbar T / md, \qquad (9)$$

where $m$ is the mass of the atom. The prediction for the numerically calculated split distance agrees well asymptotically with the obtained fringe spacings for the case of sufficiently large distance splitting (dashed line). Inconsistency between the prediction and the obtained data for smaller distance splitting can be ascribed to the repulsive interaction in the BECs [9]. Despite this inconsistency, from the asymptotical behavior and a numerical calculation with the experimental conditions, we were able to estimate the smallest split distance as ~2.8 μm.

### 2. Large distance splitting: positive detune

It is also possible to increase the distance between the two potential wells by sweeping up the frequency of the rf-field after the amplitude ramp is completed. In the case of large positive detuning $\hbar\omega \gg g_F\mu_B B_I$, the resonance term determines the position of the new minima dominantly. Though the bare state eigenenergies resonate with the rf-field at these positions, the coupling term acts as an effective Ioffe field for the rf-dressed potential. For this reason, the rf-field amplitude should be large enough to avoid spin flip loss at the trap minima before increasing the rf-frequency.

To realize large splitting distances, the frequency of the rf-field was raised from the initial level of 510 kHz to its final value (up to 3.84 MHz in our experiment) within 10 ms after the amplitude ramp is completed. The largely positive detuned final frequency allows us to separate the condensates sufficiently to resolve the separated BECs by taking in-situ absorption images. The static single-well was deformed into a horizontal double-well with $\delta = \pi$ and in-situ images were obtained with the axial imaging light. Fig. 6(a) shows the separated BECs for different final rf-frequencies. The measured trap separations increase linearly with the rf-frequency and are consistent with the theoretical prediction of Eq. (7), for the experimental conditions ($G \simeq 19$ T/m) [Fig. 6(b)]. We were able to split BECs over distances of up to ~57 μm without significant loss or heating.

### 3. Dependency on the polarization of the rf-field

The coupling term of the rf-induced double-well potential only depends on the linearly polarized rf-field perpendicular to the local static field, which defines the local quantization axis (see Eq. (3)). The double-well always forms in the direction along which the transverse quadrupole field is parallel to the total rf-field. Consequently, the orientation of the split is a function of the polarization of the rf-field, Eq. (8). In our two rf-wire configuration, the polarization is determined by the ratio of the rf-field amplitudes $B_{rf}^B / B_{rf}^A$; that is, the orientation of the rf-induced double-well can be controlled as a function of the amplitude ratio of the rf-currents through the rf-wires $I_{rf}^B / I_{rf}^A$.

Fig. 7 shows the split BECs with different polarizations of the rf-field. For direct observation of the split orientation, a large distance splitting regime is applied. The frequency of the rf-field is swept linearly to 2.0 MHz at the end of the amplitude ramp and in-situ images are obtained with the axial imaging light. To show the dependency, typical polarizations of the rf-field using typical wire configurations are implemented ((a) $B_{rf}^A = B_{rf}^B$ with $\delta = \pi$, (b) $B_{rf}^A \neq 0$ and $B_{rf}^B = 0$, (c) $B_{rf}^A = B_{rf}^B$ with $\delta = 0$, (d) $B_{rf}^A = 0$ and $B_{rf}^B \neq 0$) [32]. Fig. 7 shows apparent dependence of the split orientation of the BECs on the polarization of the rf-field. We can see here that the split orientation of the matter-wave can be controlled with the relative amplitude of the rf-currents in the two rf-wires.

## IV. CONCLUSION

We construct a matter-wave beam splitter on an external atom chip integrated on a compact, transportable BEC manipulating system. The system occupies a total volume of ~0.5 m$^3$ and operates at a repetition rate of ~0.2 Hz using a maximum electric power of 765 W. By dressing rf-fields on magnetically trapped atoms, a single-well deformed into a double-well and a BEC separated into two clouds. The distance between the separated BECs was modulated as a function of the rf-field's amplitude and frequency. In addition, the orientation of the split was controlled with polarization of the rf-field. Due to the compactness, low energy consumption, and high repetition rate of the system, the beam splitter should be an ideal platform for utilizing atom interferometry based on BECs for portable applications such as high-precision measurement in space [33].

## ACKNOWLEDGMENTS

We are grateful to Prof. Dana Z. Anderson in JILA for manufacturing the atom chip and the vacuum chamber. This research was supported by a grant to the Atomic Interferometer Research Laboratory for the National Defense funded by DAPA/ADD and by Basic Science Research Program through the National Research Foundation of Korea (NRF) funded by the Ministry of Science, ICT & Future Planning (2014R1A2A2A01007460).

Table 1. Timing sequence of a typical BEC preparation cycle for matter-wave interferometry. The total cycle time is < 5 s.

| Operation | Duration |
|---|---|
| MOT | 1.2 s |
| Compressed MOT | 20 ms |
| Optical molasses | 6.2 ms |
| Optical pumping | 1 ms |
| Initial external Z-coil trapping | 2 ms |
| Ramping up the trap | 400 ms |
| Transport to atom chip trap | 100 ms |
| Compression | 100 ms |
| $1^{st}$ evaporative cooling | 2 s |
| Decompression | 100 ms |
| $2^{nd}$ evaporative cooling | 500 ms |

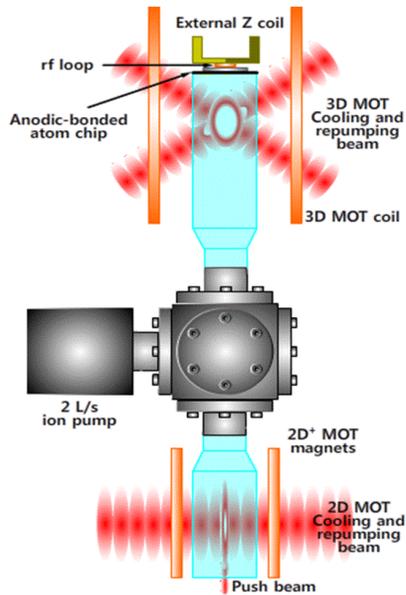

Fig. 1. Diagram of the vacuum system with an anodic bonded atom chip for a matter-wave interferometer. The vacuum system consists of two separated MOT cells to achieve a high repetition rate and a low pressure. An ultrahigh vacuum state is maintained by a 2 l/s ion pump. In the lower cell, atoms are loaded into a $2D^+$ MOT and pushed upward to supply pre-cooled $^{87}$Rb atoms. The entire experimental procedure is done in the upper cell with the transferred atoms. The distance from the bottom of the $2D^+$ MOT cell to the top of the atom chip is ~25 cm.

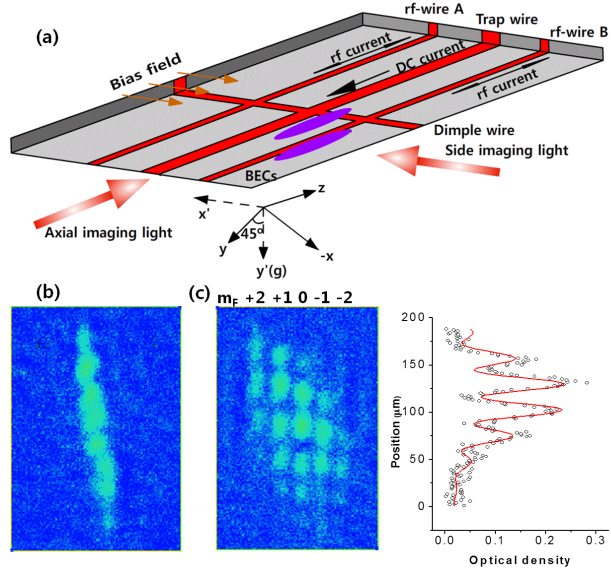

Fig. 2. Schematic of the atom chip matter-wave beam splitter. (a) Atoms were confined radially by the combined magnetic potential of a DC current carrying trap wire and an external bias field. The axial confinement is provided by a pair of end caps of the trap wire (not shown) and is enhanced by the DC current carrying dimple wire. By dressing the atoms with two oscillating rf-fields from two rf-wires beside the trap wire, the single-well is deformed into a double-well within 15 ms. (b) Typical absorption image with side imaging light after 15 ms time-of-flight expansion of two BECs split along the vertical direction. Overlapping of the interference patterns from different $m_F$ states due to spin-flip during turn-off the trapping potential disturbs the analysis of the interference patterns. (c) Left: By applying a magnetic gradient field along the z-direction during the expansion, the matter-wave interference patterns of each $m_F$ state are obtained clearly. Right: A cosine function with a Gaussian envelope is fitted into the profiles derived from the image of the $m_F = 0$ component to analyze the fringe.

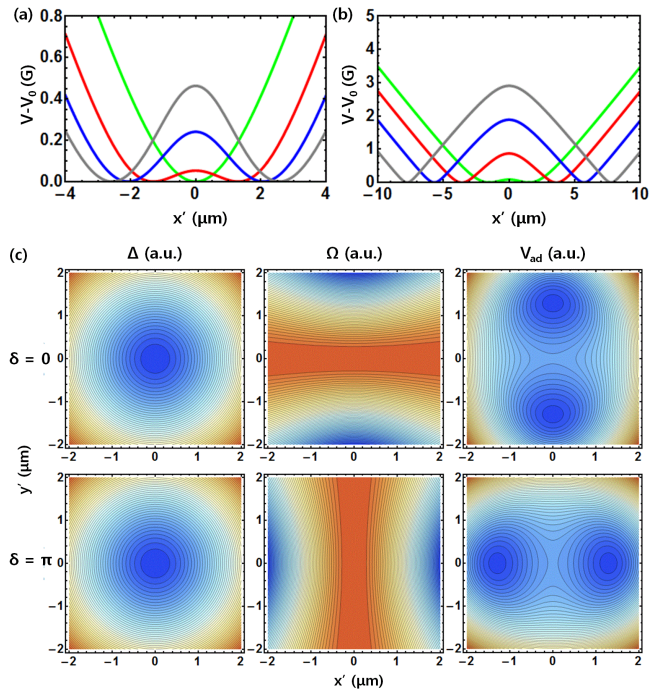

Fig. 3. rf-induced double-well potentials with different conditions: calculated for the typical set of experimental parameters $G = 50$ T/m and $B_I = 1$ G for $^{87}$Rb atoms in the $F = m_F = 2$ state. (a) Double-well potentials for different rf-frequencies $\omega$ ($B_{rf} = 1$ G): green -

$\left|g_F\mu_B B_I\right|/\hbar$, red - $2\left|g_F\mu_B B_I\right|/\hbar$, grey - $4\left|g_F\mu_B B_I\right|/\hbar$. (b) Double-well potentials for different rf-amplitudes ($\omega = 0.9\left|g_F\mu_B B_I\right|/\hbar$): green - 0 G, red - 1 G, blue - 1.8 G, grey - 2.5 G. (c) Double-well potentials for different polarizations of the rf-field.

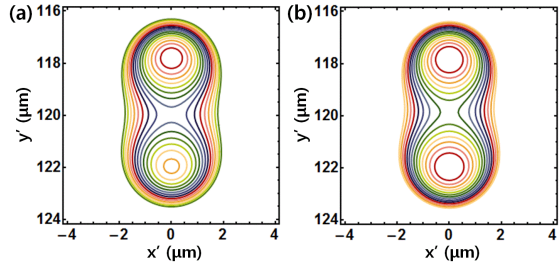

Fig. 4. Vertical double-well potentials with the condition of the interference experiments, typically $B_{rf}$ = 75 mA (4 kHz contours). (a) Asymmetric double-well potential due to gravity and inhomogeneous coupling strength. (b) The asymmetry is nearly compensated with the additional rf-field along the z-direction by the rf-current through the dimple wire, ~1.75 mA.

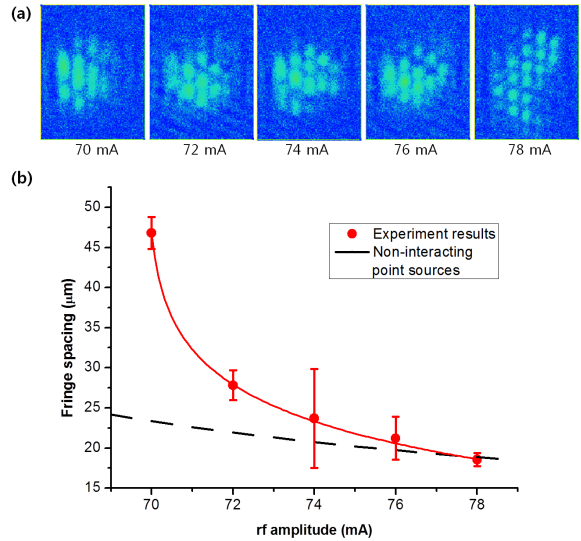

Fig. 5. (a) Matter-wave interference patterns formed after 15 ms time-of-flight expansion of the two split BECs with different rf-amplitudes. (b) The fringe spacings as a function of the rf-amplitudes (realizing different double-well separations). The dotted line indicates the expected fringe spacing based on expansion of a non-interacting gas from two points located at the two double-well minima (numerically calculated based on our experimental parameters). As the trap separation increased, the simple prediction is consistent with the obtained data. The discrepancy for small rf-amplitudes is due to interaction between the atoms. The solid line is a guide to the eye.

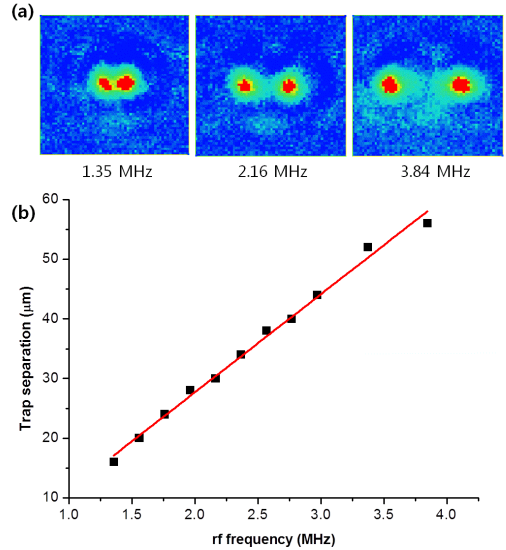

Fig. 6. Double-well trap separations for different rf-frequencies. By ramping up the rf-frequency over the Larmor frequency of the atoms at the static trap minima (~1 G corresponding to ~700 kHz) after the rf-amplitude ramp is completed, the separation increases linearly along with the rf-frequency. (a) In-situ absorption images of horizontally split BECs for different rf-frequencies. (b) The trap separations as a function of the final rf-frequency. Each absorption image has 360 × 360 μm$^2$ size.

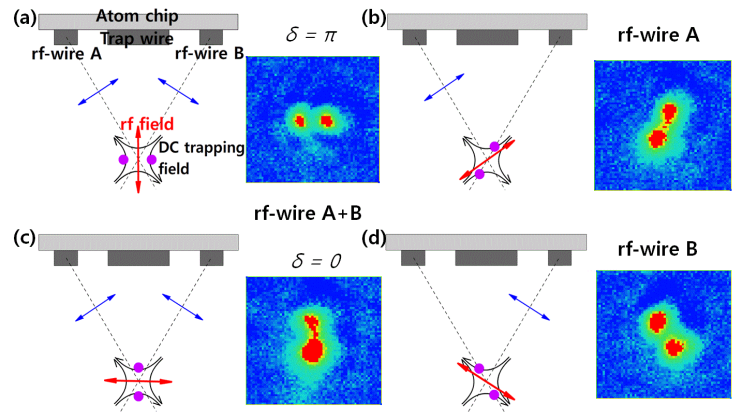

Fig. 7. Orientation of the double-well potentials for different polarizations of the rf-field. The spatial orientation of the rf-induced double-well is determined by the relative orientation of the static and the oscillating magnetic field. The black arrows and the red arrow represent the static quadrupole field in the vicinity of the trap minima and the polarization of the total rf-field, respectively. Polarization of the rf-field can be controlled with the ratio of the amplitude of the rf-fields from the two rf-wires. In-situ images with the typical polarizations: rf-wire A and B are in use with the relative phase (a) $\pi$ and (c) 0. Sole rf-wire (b) A and (d) B is in use. For clarity, two condensates are split by ~40 μm by ramping up the rf-frequency.